\documentclass[intlimits,twoside,a4paper]{article}

\usepackage{eurosym}
\usepackage{graphicx}
\usepackage{xcolor}
\usepackage{ulem}
\usepackage[cp1251]{inputenc}

\usepackage[eqsecnum]{cmpj3}

\issue{2023}{26}{3}{33603}
\doinumber{10.5488/CMP.26.33603}

\title[On the existence of a second branch]
{On the existence of a second branch of transverse collective excitations in liquid metals\thanks{The authors are grateful to Prof. Taras Bryk for numerous fruitful and stimulating discussions about this subject.}
}
\author[J.-F. Wax, N. Jakse]{J.-F. Wax\orcid{0000-0003-2209-8427}\refaddr{label1}\thanks{Corresponding author: \email{jean-francois.wax@univ-lorraine.fr}.}, N. Jakse\orcid{0000-0002-4031-0965}\refaddr{label2}}
\addresses{
\addr{label1} Laboratoire de Chimie et de Physique A2MC, Universit\'e de Lorraine, Metz, 1, boulevard Arago 57078 Metz Cedex 3, France
\addr{label2} Universit\'e Grenoble Alpes, CNRS, Grenoble INP, SIMaP, F-38000 Grenoble, France
}


\date{Received March 22, 2023, in final form April 26, 2023}
\begin{document}

\maketitle

\begin{abstract}
It was found recently that the liquid dynamics of several metals (Li, Zn, Ni, Fe, Tl, Pb) under pressure is characterized by transverse spectral functions containing an additional high-frequency peak.
To rationalize the pressure dependence of the contributions from different propagating processes to transverse spectral functions in liquid metals, ab initio molecular dynamics simulations were performed for two typical liquid metals (Na and Al) in a wide range of pressures.
The influence of density/pressure is investigated for Na by considering four pressures ranging from 15 to 147 GPa, while the temperature influence is considered for Al between 600~K in the deep supercooled liquid up to 1700~K well above the melting point at ambient pressure.
Both temperature and density dependence of the spectra of collective excitations are analyzed with a focus on the appearance of a second high-frequency mode in the transverse spectra.
A correspondence between spectra of transverse collective excitations and the peak positions of the Fourier-spectra of velocity autocorrelation functions (vibrational density of states) is found.
\printkeywords
\end{abstract}


\section{Introduction}
Prompted by experimental improvements and growing computational power, the dynamic structure of liquid metals aroused a growing interest for a couple of decades.
{Unexpected behaviours were discovered, among which a phenomenon called initially ``fast sound'', first predicted by simulations \cite{Bosse1986} and confirmed experimentally in several binary alloys with large mass ratio, related to the appearance of two dispersion branches in binary mixtures, a positive dispersion of the longitudinal acoustic excitations, and a suspected coupling between longitudinal and transverse excitations \cite{Hos2009}, still in debate.} 
Thanks to the numerical simulations, enhanced by the possibility of performing first-principles based molecular dynamics \cite{Hafner2008}, the description of atomic movements has now reached an unprecedented level of accuracy, providing quantities out of reach of experiment such as the partial dynamic structure in alloys or transverse current-correlation functions.

Recently, the existence of a second branch of transverse excitations was revealed {by numerical simulations} under high pressure in molten Li~\cite{Bry2015}, Na~\cite{Mar2016}, Fe~\cite{Mar2016b}, Zn~\cite{Rio2017}, Al~\cite{Jak2019}, Pb~\cite{Bry2019}, but also under ambient pressure in Tl~\cite{Bry2018}, In~\cite{Bry2020}, and also possibly in some others~\cite{Rio2020}, but not in Na~\cite{Bry2016, Bry2016a}.
{This phenomenon cannot be observed experimentally because only numerical simulations allow one to get these properties in liquids.}
It should also be noted that an hydrodynamic propagation gap exists since transverse excitations do not propagate in liquids over macroscopic distances. 
More importantly, molecular dynamic simulations revealed, concomitantly to the emergence of this second transverse branch, the appearance of a second peak in the spectral density of the velocity autocorrelation function (VACF) (or vibrational density of states)~\cite{Bry2018,Bry2019,Jak2019}.

The aim of the present work is to give a deeper and comprehensive view of this second branch of transverse excitations.
For this purpose, ab initio molecular dynamics simulations (AIMD) are performed since transverse excitations cannot be measured experimentally in liquids.
Moreover, AIMD provide a more accurate description of the interactions under high pressures, the thermodynamic conditions where these transverse modes are believed to appear.
Therefore, liquid sodium at high pressure ranging from 15 to 147~GPa is first studied in order to investigate the effect of this parameter on the appearance of two transverse dispersion branches, which were not considered in our previous work \cite{Bry2020}.
In a second step, molten Al is studied at ambient pressure and over a wide temperature interval, from the undercooled state (at 600~K) up to 1700~K, well-above the melting point (933~K).
This will allow us to highlight this phenomenon as a function of temperature and especially if it survives below the melting point.
This will complement the study of  \cite{Jak2019} since we  investigate the ambient pressure, but above all we investigate the undercooled state.
As we will see, our results show that these two dispersion branches, as well as the corresponding peaks of the spectral density of the VACF, probably always exist, but they mostly overlap under ambient conditions in most cases.

{In what follows, we  first describe the theoretical background in section~\ref{sec2}, including the simulation parameters, the state points investigated and the structural functions considered.
Then, in section~\ref{sec3}, we  present our results, starting from sodium where we  mainly focus on the density (pressure) dependence of the transverse excitations.
The temperature influence will be discussed looking at Al.
Finally, we will summarize our results in section~\ref{sec4}.}

\section{Theoretical background}\label{sec2}
\subsection{Ab initio molecular dynamics simulations}
In order to make sure that the description of the interactions in the metals is accurate whichever their density and temperature, ab initio molecular dynamics simulations were performed with Vienna Ab initio Simulation Package (VASP)~\cite{Kresse1996}.
We simulated an assembly of $N=300$ atoms for sodium and of $N=256$ atoms for aluminum.
7 electrons per atom were taken into account for the description of the electronic structure in the case of sodium, and only 3 electrons in the case of aluminum since the pressure ranges and so the core-overlaps were different. 
The volume $V=L^3$ of the cubic box of side $L$ was chosen to reproduce the desired density and the temperature $T$ controlled using Nos\'e thermostat.
The simulations were, therefore, carried out in the NVT ensemble rather than in NPT since the size of the simulation box governs the obtainable $q$-values which are of importance for the functions considered.
For instance, the lowest reachable $q$-value is $q_{\text{min}} = 2 \piup /  L$.
Any fluctuation of the box size induced by constant pressure constraint would lead to a fluctuation of the $q$-values at which the dynamic structure functions are evaluated.

Electron-ion interactions were described using PAW potentials \cite{Kresse1999} and PBE GGA \cite{Perdew1996} exchange correlation description for Na and LDA \cite{Ceperley1980,Perdew1981} for Al since the density ranges were different.
This is also the reason why we considered 7 electrons per atom for Na and only 3 electrons in the case of Al.
More importantly, it was shown that LDA gives an accurate description of the dynamics of liquid Al \cite{Jakse2013,Demmel2021}.
The time steps were 3 fs and 1.5 fs long for Na and Al, respectively.
The production stage lasted 10 000 steps for Na and 40 000 for Al.

\subsection{Simulated state points}
\begin{figure}[tbp]
\centerline{
\includegraphics[width=.48\textwidth]{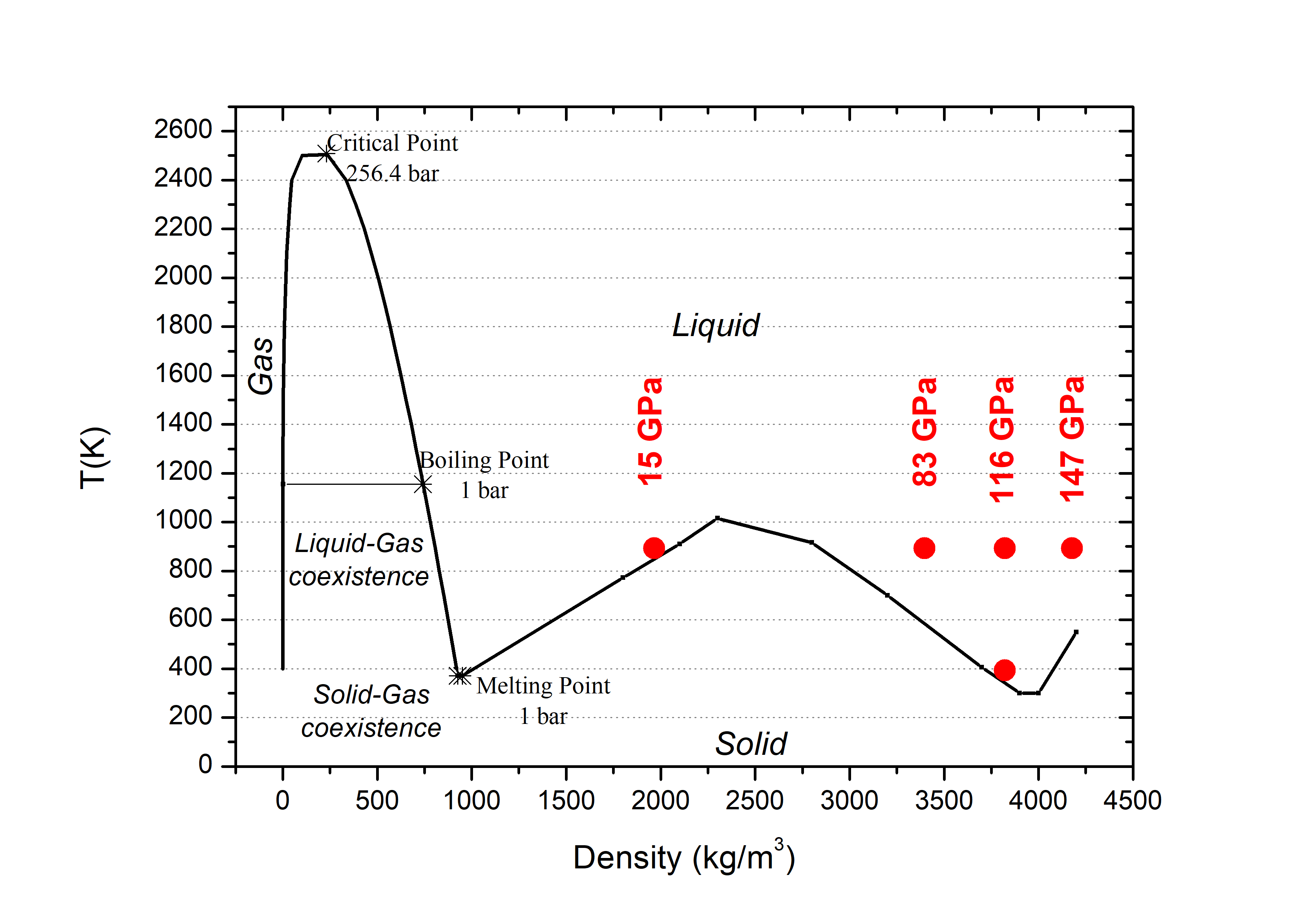}
\includegraphics[width=.48\textwidth]{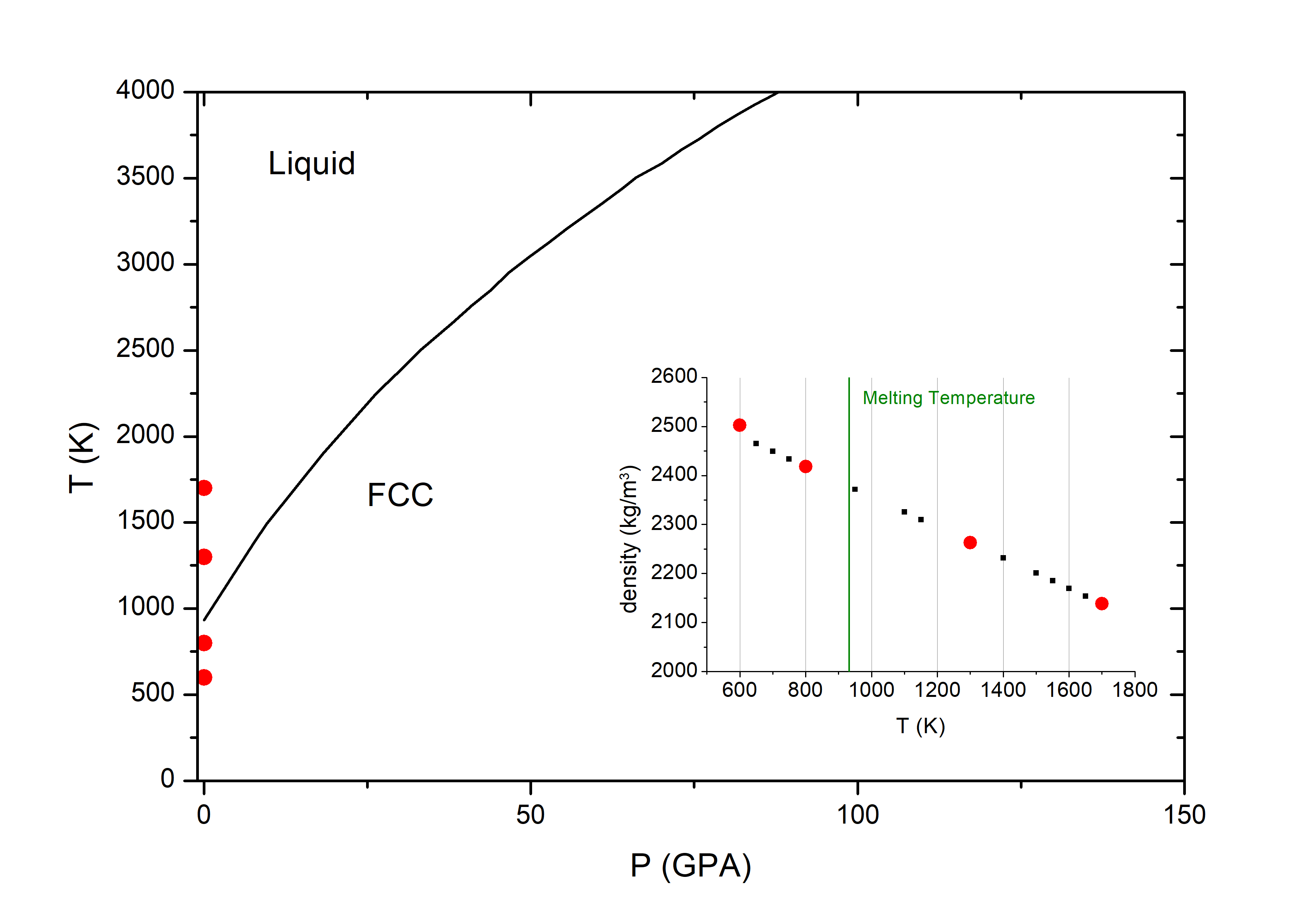}
}
\caption{{(Colour online) Location of the simulated state points of Na (left-hand) and Al (right-hand) in their phase diagrams (according to \cite{Meyer2016} for Na and to \cite{Jak2019} for Al). For Al, the inset displays the density versus temperature.}}
\label{fig1}
\end{figure}

The investigated state points for Na and Al are presented in figure \ref{fig1}.
In the case of Na, they are superimposed on the phase diagram, while in the case of Al, they are plotted along the ambient pressure isobar.
The corresponding temperatures and densities, as well as the pressure values computed from our simulations in the case of Na are gathered in table \ref{tab1}.
In the case of Na, we investigated the pressure influence along the 893~K isotherm, as well as the temperature influence along the 116~GPa isobar.
As for Al, we studied the temperature influence at ambient pressure over the melting point (1300 and 1700~K) and in the undercooled regime (600 and 800~K).

\begin{table}[!t]
\caption{Thermodynamic state points for the simulations of Na and Al.
	In the case of Al, they are performed at ambient pressure.
	The position $q_p$ of the first peak of $S(q)$ is also indicated. 
}
\vspace{2ex}
\centerline{
	\begin{tabular}{cccccc}
		\hline\hline
		&  & Na &  &  &  \\ \hline
		$T$(K) & 893 & 893 & 393 & 893 & 893 \\
		P(GPa) & 15 & 83 & 116 & 116 & 147 \\
		$\rho $(\AA $^{-3}$) & 0.0514 & 0.0889 & 0.1000 & 0.1000 & 0.1093 \\
		$q_p $(\AA$^{-1}$ ) & 2.55 & 3.07 & 3.19 & 3.19 & 3.29 \\ \hline
		&  & Al &  &  &  \\ \hline
		$T$(K) & $600$ & $800$ & $1300$ & $1700$ &  \\
		$\rho $(\AA $^{-3}$) & 0.0558 & 0.0540 & 0.0505 & 0.0477 &  \\
		$q_p $(\AA$^{-1}$ ) & 2.76 & 2.75 & 2.74 & 2.69 & \\
		\hline\hline
	\end{tabular}%
 }
\label{tab1}
\end{table} 

\subsection{Dynamic structure functions}
Many functions are available to describe different aspects of the dynamic structure of a liquid.
In this study, we are interested in the transverse excitations and their dispersion.
Therefore, we computed the transverse current correlation function
\begin{equation}
C_T ({\mathbf q}, t) = \frac{1}{2N} \left\langle \mathbf{j}_T^* ({\mathbf q}, 0) \cdot \mathbf{j}_T ({\mathbf q}, t) \right\rangle .
\end{equation}
This was obtained from the transverse part of the particle current 
\begin{equation}
\mathbf{j}({\mathbf q},t) = \sum_{l=1}^{N} {\mathbf v_l} (t) \exp \left[ i{\mathbf q}\cdot{\mathbf r_l} (t) \right] = \mathbf{j}_L ({\mathbf q},t) + \mathbf{j}_T ({\mathbf q},t),
\end{equation}
which corresponds to the case where $\mathbf q$ and $\mathbf{v}_l$ are orthogonal.
In this expression, $\mathbf{r}_l$ and $\mathbf{v}_l$ are the position and velocity of particle $l$, respectively.
The $\mathbf q$-values depend on the side of the box.
We only considered wavevectors that can be written as $\mathbf q = (n_x, n_y, n_z) 2 \piup / L$, where only one of the three integers is non-zero.
Due to isotropy of the liquid, $C_T ({\mathbf q}, t) \equiv C_T (q, t)$, so that the statistical accuracy is the same for each $q$-value considered.

\begin{figure}[hbp]
\centerline{
\includegraphics[width=.48\textwidth]{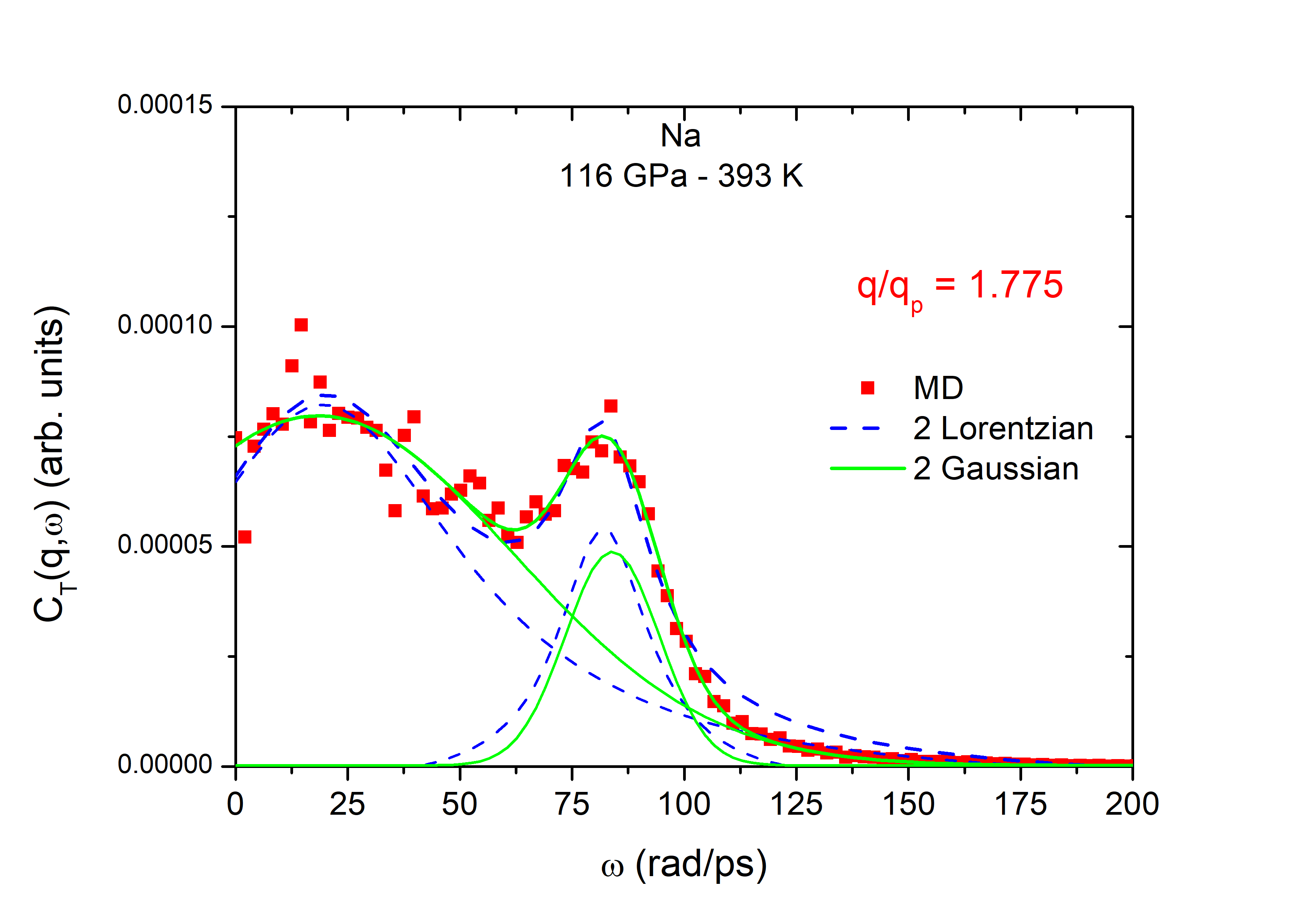}
}
\caption{(Colour online) Example of a fit of the transverse current correlation function with two Gaussian and two Lorentzian functions.}
\label{fig2}
\end{figure}

As we are interested in the propagating collective modes, we advantageously computed the spectral density of the transverse current correlation function, $C_T ({\mathbf q}, \omega )$, since such modes appear as peaks in this function.
In order to locate their frequencies, we fit the results obtained at a given $q$-value with a sum of two Gaussian functions.
Although this choice could be discussed, it appears to be efficient and in agreement with the observed line-shape, within the statistical uncertainty as can be seen in figure \ref{fig2}.
In this figure, we also plot a fit with two Lorentzian functions for the sake of comparison.
Although this last expression would be better grounded on a fundamental point of view, for some reason, Gaussian functions better reproduce the large-frequency side of the peaks as can be seen.
Anyway, differences between frequency values obtained from both expressions are small.

We also computed the spectral density $\psi (\omega )$ of the velocity autocorrelation function (VACF) 
\begin{equation}
\psi (t) =\frac{1}{N} \left\langle \mathbf{v_l}(0) \cdot \mathbf{v_l} (t) \right\rangle .
\end{equation}
This will allow us to compare the frequencies involved in these functions to the dispersion curves.

\section{Results}\label{sec3}
\subsection{Sodium}
\begin{figure}[tbp]
\centerline{
\includegraphics[width=.48\textwidth]{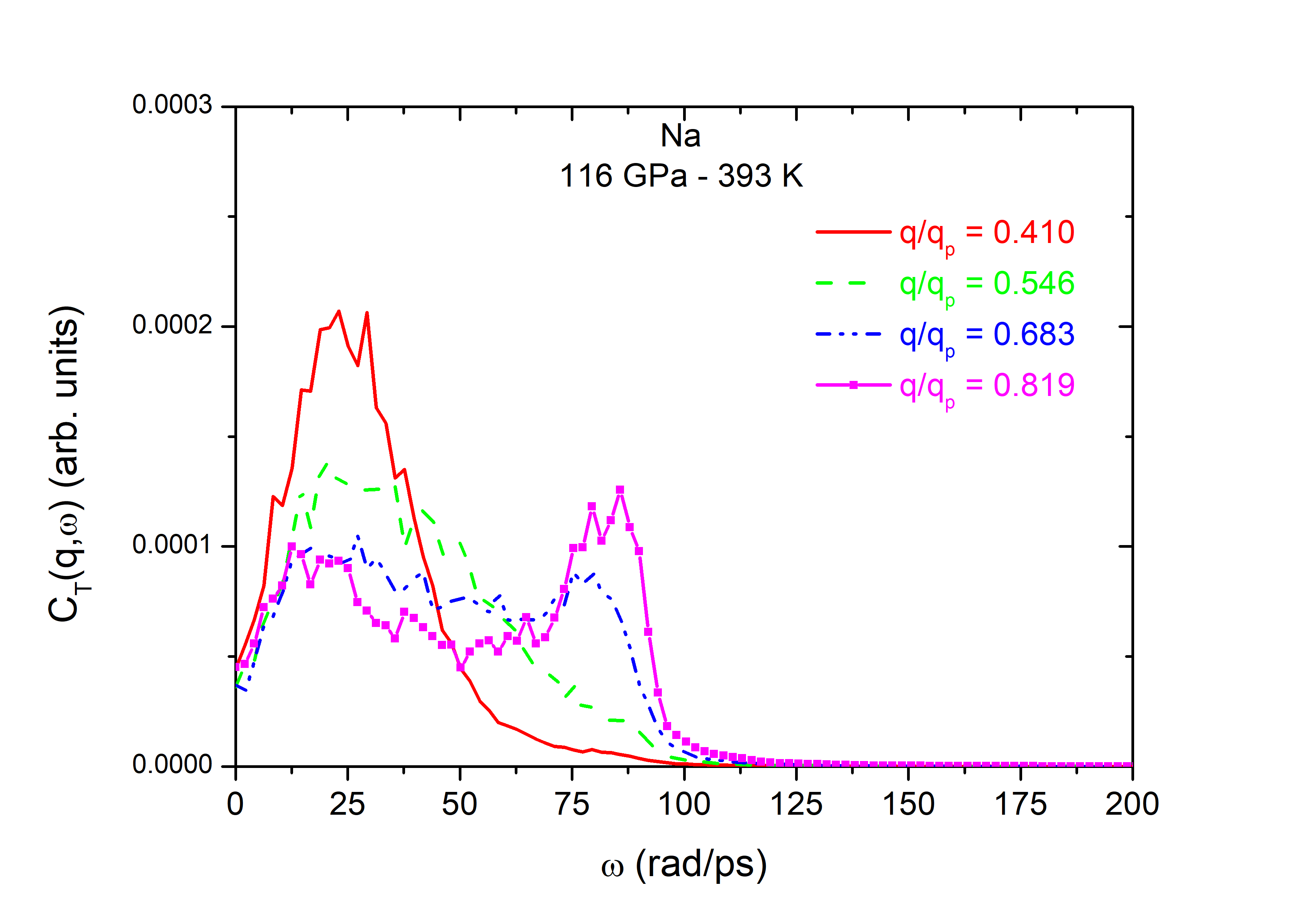}
\includegraphics[width=.48\textwidth]{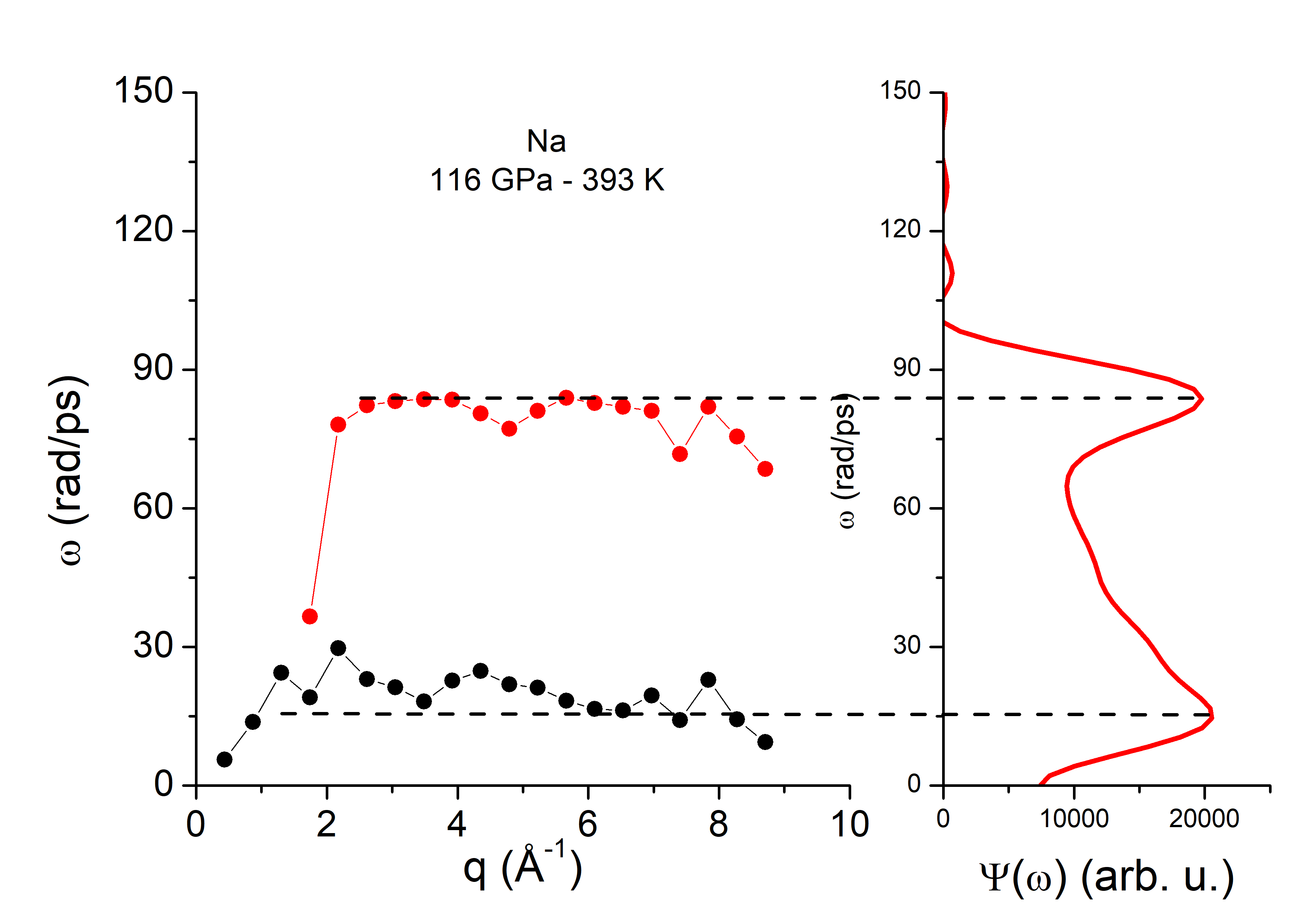}
}
\caption{(Colour online) Transverse excitations of Na at 116 GPa and 393 K.
Spectral density (left-hand); dispersion branches (centre) compared with the spectral density of the VACF (right-hand).
Lines are guides for the eyes.
}
\label{Fig3}
\end{figure}

We first consider the state point (116 GPa, 393 K).
The spectral density of the transverse excitations is plotted at some selected $q$-values in figure \ref{Fig3}.
Transverse collective excitations are clearly visible in these curves as peaks at non-zero frequencies.
In fact, the propagation gap that does exist at small $q$ cannot be seen in our results since the lowest reachable value $q_{\text{min}} = 2 \piup / L$ is too high.
The second mode is nicely visible at high $q$.
In fact, a second peak seems to appear for $q$-values higher than $q_p /2$ and is still present at the highest value we considered, about $3q_p$ (not shown in figure \ref{Fig3}).

Using the fitting procedure described in the previous section, we are able to estimate the peak positions and plot the dispersion branches of the transverse excitations.
These are displayed in figure \ref{Fig3}.
The high frequency mode is not observed below $q_p /2$ or, at least, cannot be resolved from our data.
It quickly reaches a plateau and does not show any dispersion within the limits of our determination procedure.
This is not the case of the low
frequency mode that starts from $\omega=0$ at a finite $q$-value smaller than $q_{\text{min}}$, reaches a maximum at about 2.2 \AA$^{-1}$ and slightly
oscillates around a finite value at higher $q$.

The dispersion branches are compared with the spectral density of the VACF.
The non-zero low-frequency limit related to the self-diffusion is recovered indicating that the metal is undoubtedly liquid.
However, at this state point, the function displays two peaks while only one is observed in sodium at ambient pressure~\cite{Wax2001}.
These peaks at non-zero frequencies correspond to vibrations of the atoms in their neighbours cage.
Probably, this could be related to some anisotropy of the local environment of the atom.
{Interestingly, this double peaked spectral density is also observed in simulations of high density sodium performed using a spherically symmetric pair potential \cite{Wax2023}.
As a consequence, it should not be the result of angle dependent interactions such as three-body effects, but rather is caused by local topological constraints that create vibration anisotropy in the cage of neighbouring atoms.}

Very interesting is the coincidence of the positions of these peaks  and the frequencies in the collective excitations.
This is all the more noticeable since VACF is an individual property while the dynamic structure is a collective one.
{This is probably related to the self-part contained in $C_T ({\mathbf q}, t)$ corresponding to the correlation of a particle with itself.
This would deserve further investigations.}

\begin{figure}[!t]
\centerline{
\includegraphics[width=.48\textwidth]{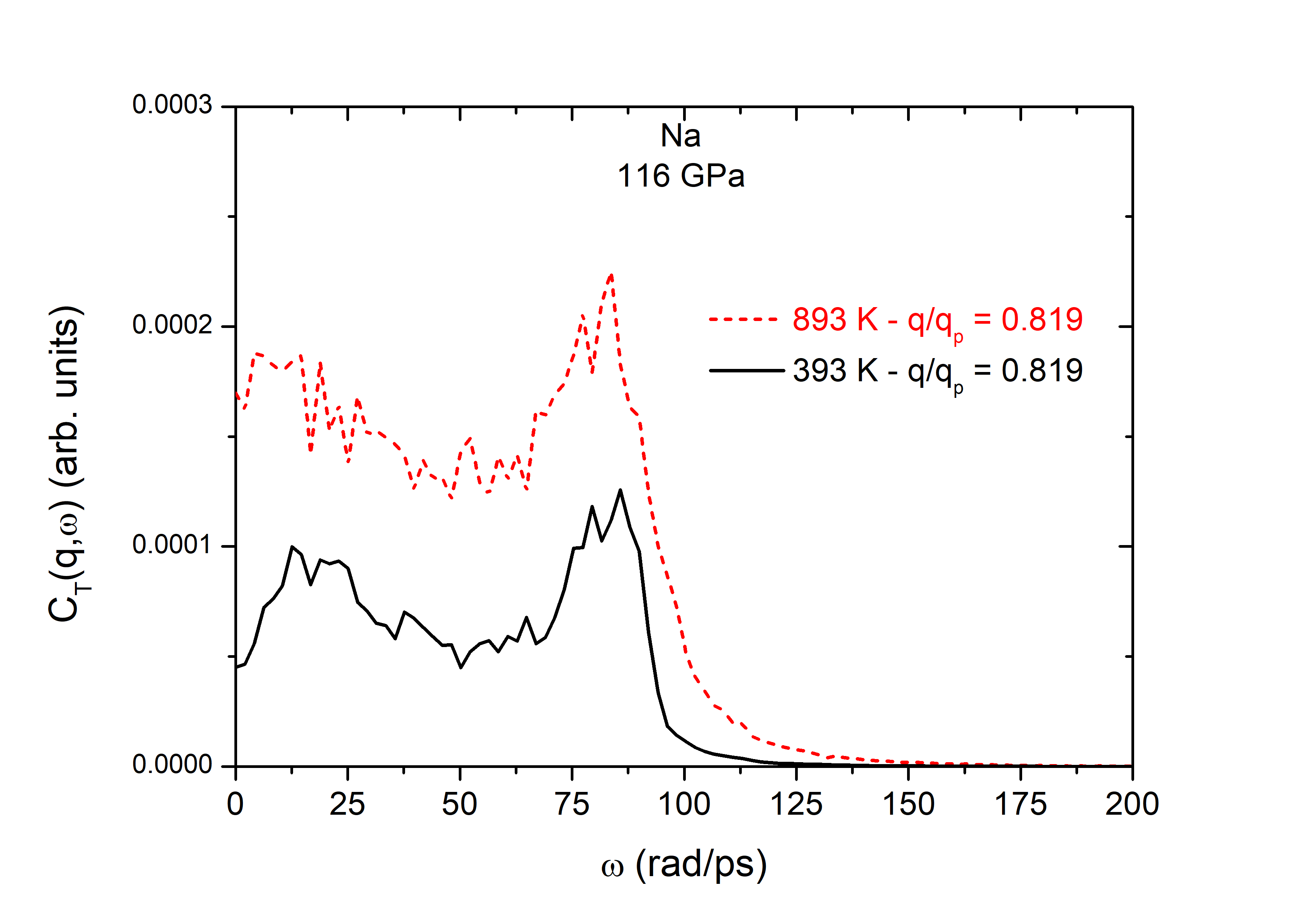}
\includegraphics[width=.48\textwidth]{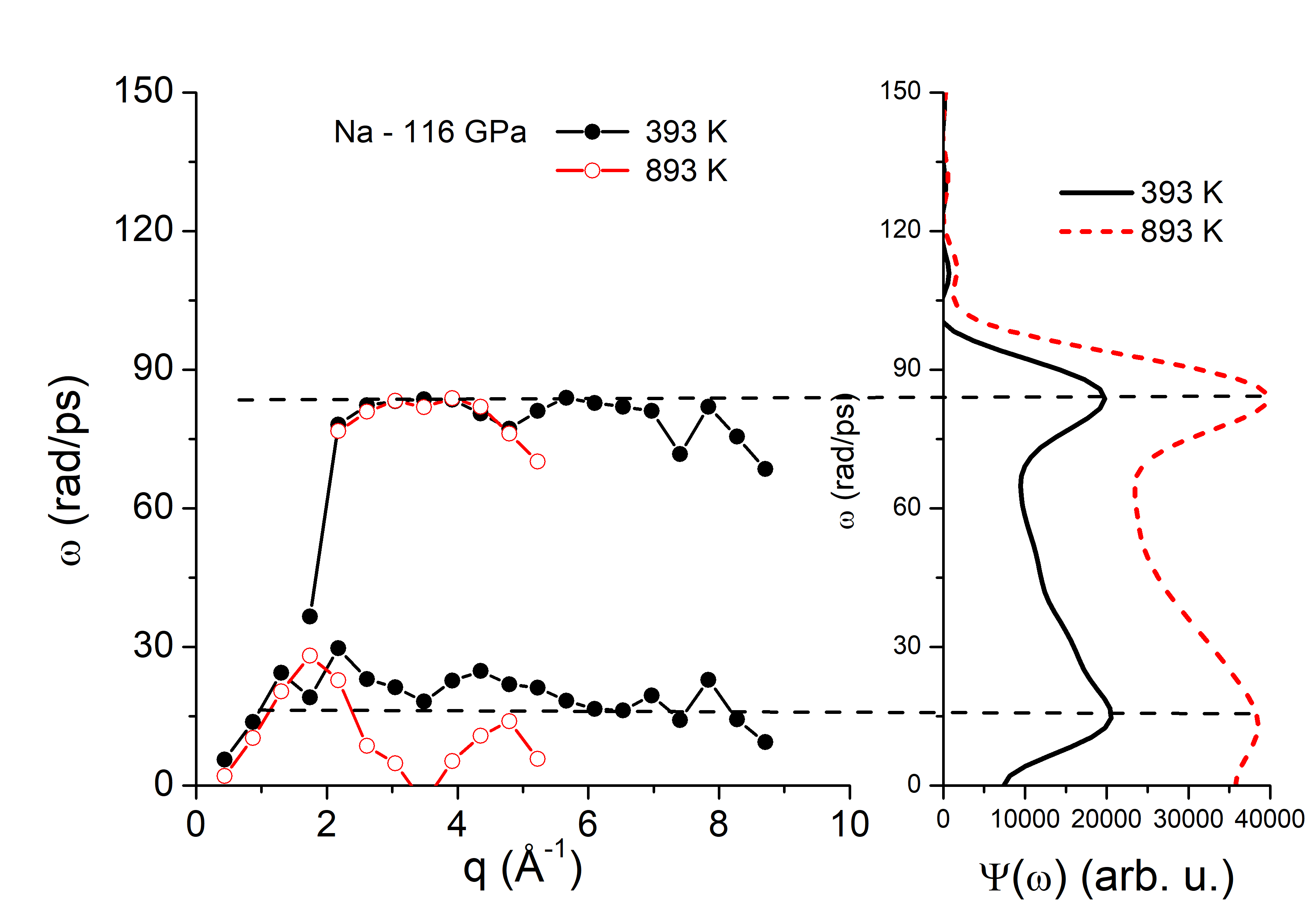}
}
\caption{(Colour online) Transverse excitations of Na at 116 GPa and two temperatures.
Spectral density at a single $q$-value (left-hand) for 373 (black solid line) and 893 K (red dashed-line); dispersion branches (centre) at 373 and 893 K, compared with the spectral density of the VACF (right-hand).
Lines are guide for the eyes.}
\label{Fig5}
\end{figure}

The influence of temperature is illustrated in figure \ref{Fig5}.
As can be seen, the peak positions are unchanged.
However, the noise gets stronger as temperature increases and the peaks become slightly broader so that the determination of the peak positions gets harder.
This is why the $q$-range in which the data were obtained is shorter.
Anyway, the two modes still exist at high $q$ even if we were not able to resolve their positions.
Considering the spectral density of the VACF, we can see that the peak position is unchanged too, so that the coincidence with the dispersion curves is still observed.
We will return to the influence of the temperature while considering the results for liquid Al.

\begin{figure}[tbp]
\centerline{
\includegraphics[width=.48\textwidth]{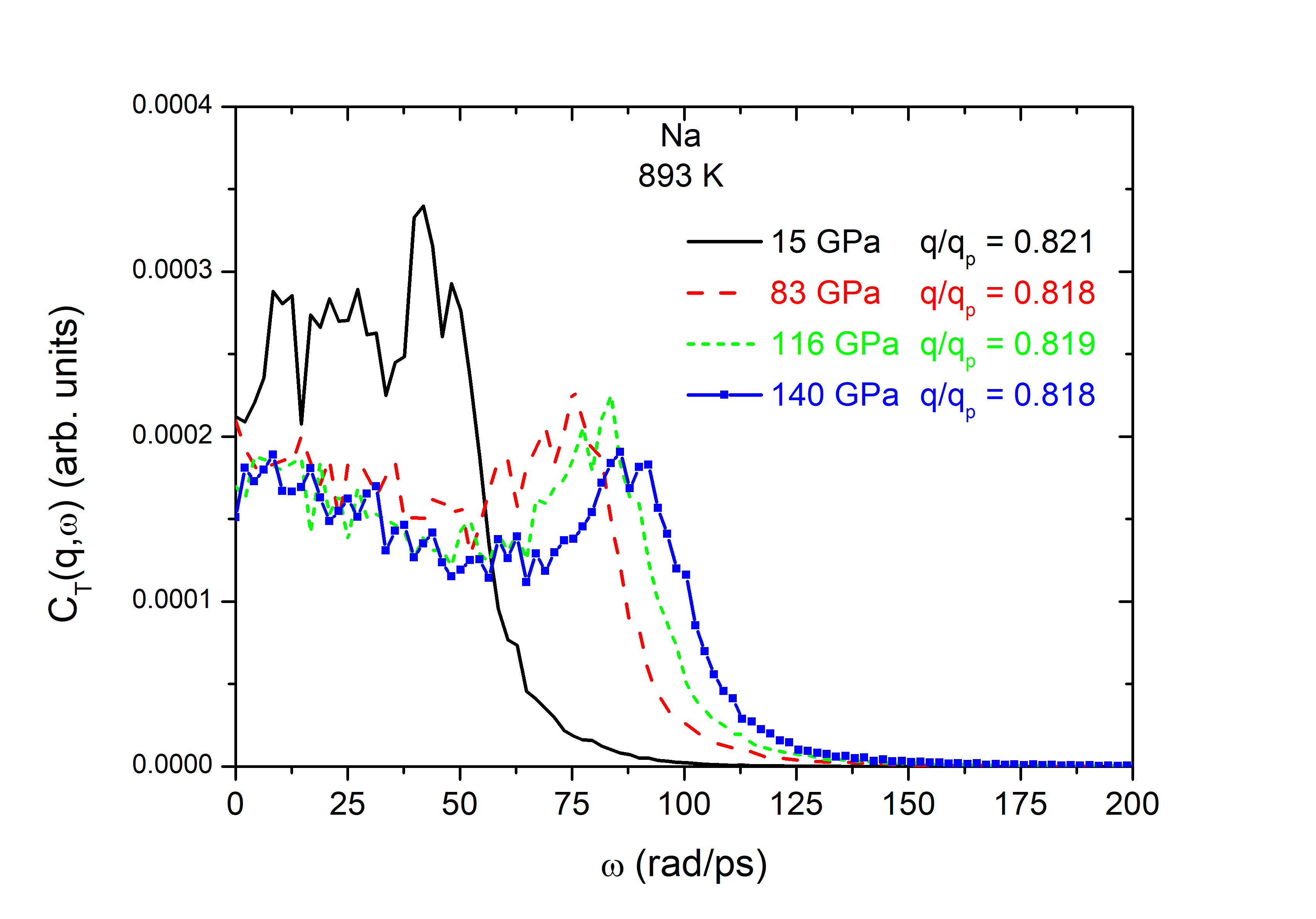}
\includegraphics[width=.48\textwidth]{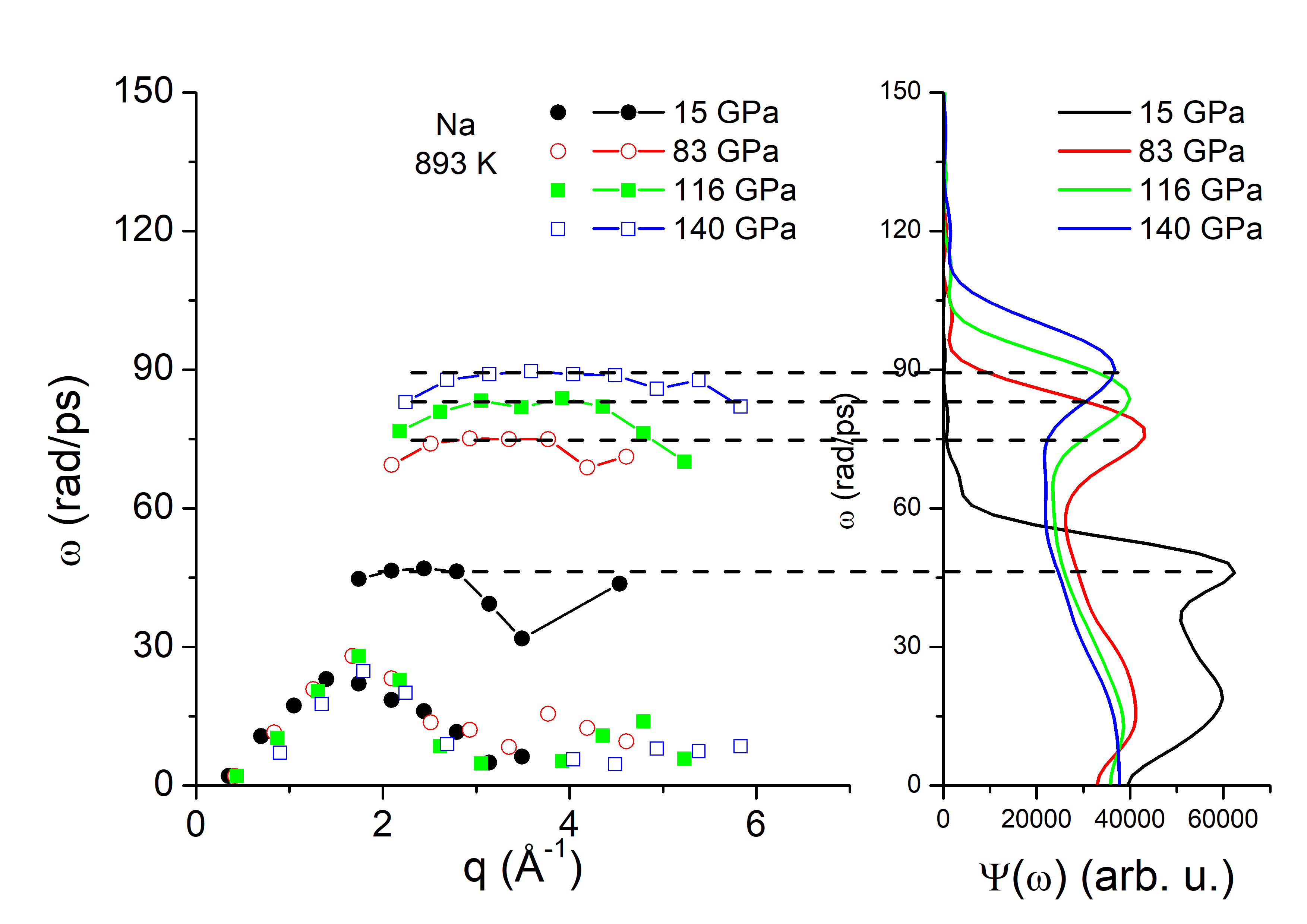}
}
\caption{(Colour online) Transverse excitations of Na at 893 K and four pressures, namely 15, 83, 115 and 140 GPa.
Spectral density at a single $q$-value (left-hand); dispersion branches (centre) compared with the spectral density of the VACF (right-hand).
Lines are guide for the eyes.}
\label{Fig7}
\end{figure}

The influence of pressure is presented in figure \ref{Fig7} which displays the results obtained at constant temperature 893~K and pressures ranging from 15 GPa to 140 GPa.
As can be seen, the two peaks always exist over the pressure range considered.
However, the high-frequency peak moves to lower frequencies as pressure decreases and becomes harder to distinguish from the low-frequency peak.
We can extrapolate that both peaks will  merge at ambient pressure as is commonly observed.
The plotted dispersion curves  nicely illustrate this trend.
We still observe that the high-frequency branch remains flat.
On the other hand, the evolution of the low-frequency branch with pressure is harder to evaluate due to the statistical uncertainty.
The very good coincidence between the dispersion branches and the VACF spectral density is noticeable, especially with regard to the high-frequency branch.
Interestingly, the linear density dependence of the high-frequency mode reported in~\cite{Bry2020} and obeyed by our results allows an extrapolation to the ambient density (851~kg/m$^3$).
In this way, we evaluate the high-frequency peak to be located at about 26~rad/ps under ambient conditions, a value in the range of the low-frequency mode.
Thus, we can claim that, in the case of sodium, both high- and low-frequency transverse modes, as well as high- and low-frequency peaks of the VACF spectrum coincide at ambient pressure, but are too close to each other so that they remained unnoticed in the former studies~\cite{Bry2016,Bry2016a}.

\subsection{Aluminum}

We turn now to the study of liquid Al.
A previous study \cite{Jak2019} at growing pressure along the melting line is consistent with our results obtained for Na and discussed in the preceding section.
Therefore, we focus here on the influence of temperature at ambient pressure over a range (600 to 1700~K) including the undercooled state (melting temperature is about 933~K).

\begin{figure}[tbp]
\centerline{
\includegraphics[width=.48\textwidth]{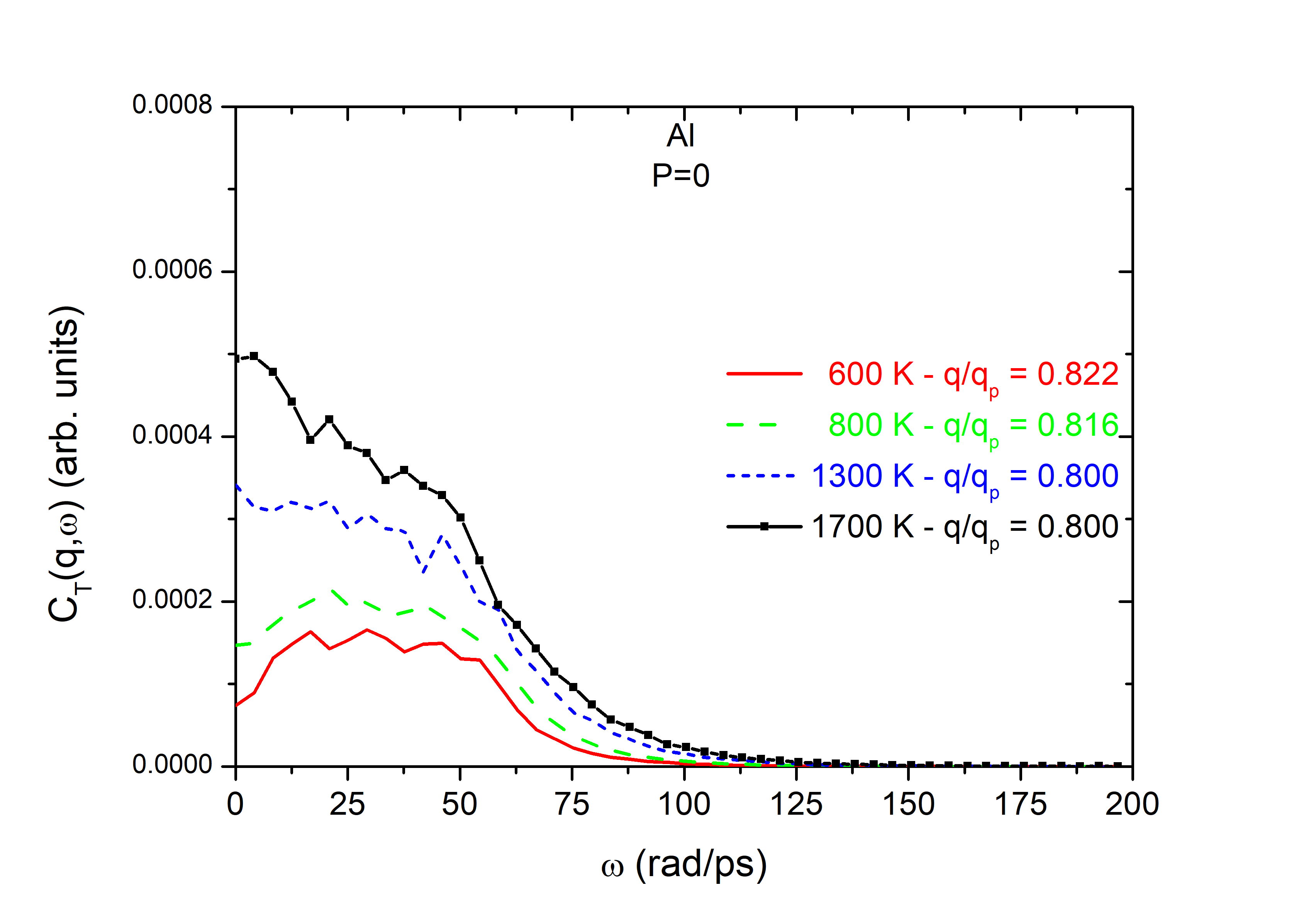}
\includegraphics[width=.48\textwidth]{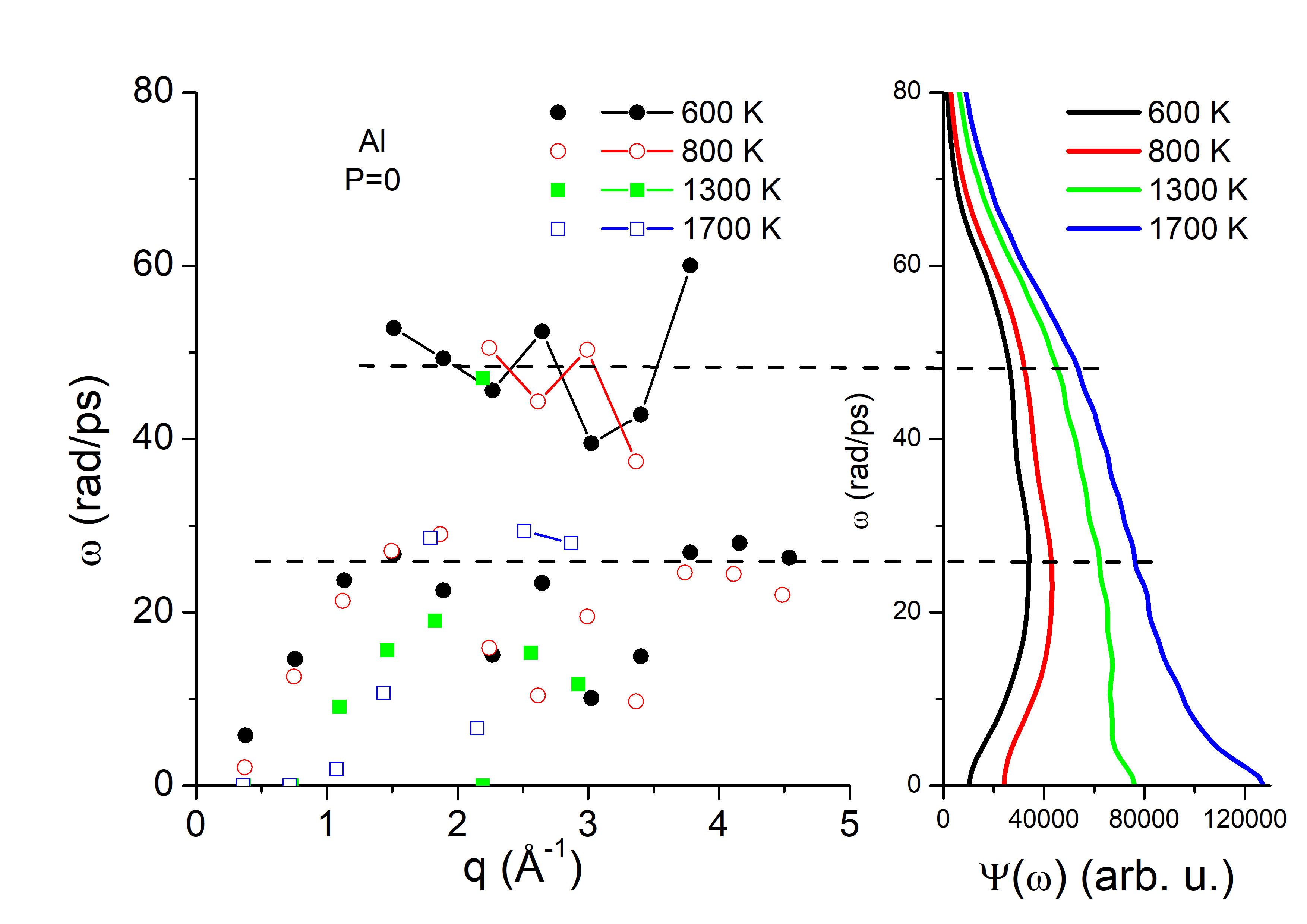}
}
\caption{(Colour online) Transverse excitations of Al at ambient pressure and four temperatures, namely 600, 800, 1300, and 1700~K.
Spectral density (left-hand); dispersion branches (centre) compared with the spectral density of the VACF (right-hand).
Lines are guides for the eyes.}
\label{Fig9}
\end{figure}

Looking at the spectral density in figure \ref{Fig9}, we can see that two peaks are still distinguishable in the shape of the spectral density of the transverse excitations.
Nevertheless, as can be expected from the trends drawn from our study of Na, at ambient pressure, both peaks are very close to each other and hard to disentangle.
However, their existence is clearly visible, especially at low temperatures.

Consequently, the dispersion branches are harder to extract accurately.
Following the same procedure as in the case of Na, we obtained the curves displayed in figure \ref{Fig9}.
Results are much more scattered, especially at high temperature.
Indeed, the low frequency branch is nicely recreated in both undercooled states, while its extraction is really problematic as temperature rises above the melting point.
This also applies to the high frequency branch, but some features can still be observed.
The first one is that it appears at $q$-values higher than $q_p /2$.
The second one is that, within the uncertainty of our results, no influence of the temperature over the frequencies of the excitations can be noticed.
Probably, a widening of the propagation gap could be suspected, but our data do not allow to conclude about this point.
Finally, a coincidence of the branches with the spectral density of the VACF is still observed.

\section{Conclusion}\label{sec4}
To summarize, in this study, we focused on the spectral density of the transverse collective excitations and its evolution with temperature and pressure.

A first conclusion to be drawn is that the existence of a second branch at $q > q_p/2$ is a general feature of liquid metals. At least is it observed here in the case of Na and Al.
As pressure increases, this high frequency branch moves away from the low frequency branch.
As temperature increases, no significant shift in the peak positions is observed but peaks broaden and both high and low frequency excitations get harder to distinguish.
This explains why this second branch can sometimes go unnoticed, either because pressure is too low so that peaks are too close or because temperature is too high so that peaks are too broad.

A second conclusion is that the coincidence between the collective transverse excitation frequencies, especially the high frequency branch, and the vibrational density of states (VDOS) ones, is rigorously observed in all the situations considered in this study.
Consequently, the second peak sometimes observed in the spectral density of the VACF can reasonably be attributed to transverse excitations.
In fact, Gaskell and Miler \cite{Gaskell1978, Gaskell1978a, Gaskell1978b} proposed that the VACF may contain both longitudinal and transverse contributions on the basis of the mode-coupling theory.
A question remains about the exact nature of these two branches and the difference between the underlying excitations.

\ukrainianpart

\title{Про існування другої гілки поперечних збуджень в рідких металах}
\author{Ж.-Ф. Вакс\refaddr{label1}, Н. Жакс\refaddr{label2}}

\addresses{
	\addr{label1} Лабораторія  хімії та фізики A2MC, Університет Лотарiнгії, Метц, 1, бульвар Араго  57078 Метц Cedex 3, Франція
	\addr{label2} Університет Гренобля, CNRS, Гренобль INP, SIMaP, F-38000 Гренобль, Франція
}

\makeukrtitle

\begin{abstract}
	Нещодавно було виявлено, що поперечні спектральні функції, якими характеризується рідинна динаміка багатьох металів  (Li, Zn, Ni, Fe, Tl, Pb) під дією високого тиску,
	містять додатковий високочастотний пік. Для вивчення залежності внесків від різних пропагаторних процесів до поперечних спектральних функцій в рідких металах під дією
	високого тиску, проведено першопринципні моделювання методом молекулярної динаміки для двох типових рідких металів (Na і Al) в широкому діапазоні 
	тисків. Вплив густини/тиску досліджено для Na при чотирьох тисках між 15 та 147 ГПа, в той час як вплив температури вивчено для Al 
	у діапазоні від 600~K у глибоко переохолодженій рідині аж до 1700~K, що суттєво вище точки плавлення при атмосферному тиску.
	Проаналізовано температурні та густинні залежності спектрів колективних збуджень; особлива увага приділялася появі другої високочастотної моди в
	поперечних спектрах. Зроб\-ле\-но висновок про кореляцію між спектрами поперечних колективних збуджень та положеннями піків у Фур'є-спектрах автокореляційних функцій швидкостей, що пов'язано з густиною вібраційних станів.
	\keywords дослідження при високих тисках, рідини, метали, першопринципна молекулярна динаміка, колективна динаміка
\end{abstract}

\lastpage
\end{document}